\documentclass[10pt,conference]{IEEEtran}
\usepackage{tikz}
\usetikzlibrary{shapes.geometric, arrows}

\ifCLASSOPTIONcompsoc
  \usepackage[nocompress]{cite}
\else
  \usepackage{cite}
\fi

\usepackage{amsmath}
\usepackage[linesnumbered,ruled]{algorithm2e}
\usepackage{algpseudocode}
\usepackage{float}
\usepackage{multirow}
\usepackage{tabularx}
\usepackage{titlesec}
\usepackage{lipsum} 

\ifCLASSOPTIONcompsoc
 \usepackage[caption=false,font=footnotesize,labelfont=sf,textfont=sf]{subfig}
\else
 \usepackage[caption=false,font=footnotesize]{subfig}
\fi

\usepackage{titlesec}
\titlespacing{\section}{0pt}{\parskip}{-\parskip}

\begin{document}

\title{
Towards Mitigating ChatGPT's Negative Impact on Education: Optimizing Question Design through Bloom's Taxonomy}

\author{Saber~Elsayed \\
School of Engineering and IT,\\
University of New South Wales Canberra
}


\IEEEtitleabstractindextext{%
\begin{abstract}

The popularity of generative text AI tools in answering questions has led to concerns regarding their potential negative impact on students' academic performance and the challenges that educators face in evaluating student learning. To address these concerns, this paper introduces an evolutionary approach that aims to identify the best set of Bloom's taxonomy keywords to generate questions that these tools have low confidence in answering. The effectiveness of this approach is evaluated through a case study that uses questions from a Data Structures and Representation course being taught at the University of New South Wales in Canberra, Australia. The results demonstrate that the optimization algorithm is able to find keywords from different cognitive levels to create questions that ChatGPT has low confidence to answer. This study is a step forward to offer valuable insights for educators seeking to create more effective questions that promote critical thinking among students.

\end{abstract}

}

\maketitle

\IEEEdisplaynontitleabstractindextext

%
\IEEEpeerreviewmaketitle

\ifCLASSOPTIONcompsoc

\IEEEraisesectionheading{\section{Introduction}\label{sec:introduction}}
\else

\section{Introduction}
\label{sec:introduction}
\fi

Generative text artificial intelligence (AI) tools, such as ChatGPT \cite{openai-chatgpt}, have become increasingly popular for answering questions and generating content. These tools use deep learning algorithms that can generate human-like responses to questions by learning from vast amounts of text data \cite{lund2023chatting}. They have also demonstrated great performance in answering programming questions, making them an attractive resource for students seeking quick answers \cite{savelka2023can}.

However, concerns have been raised about the potential negative impact of these tools on student learning and the challenges faced by educators in evaluating student understanding and knowledge \cite{baidoo2023education, savelka2023can}. Excessive reliance on AI tools for assessment can lead to students developing poor critical thinking skills and not applying what they have learned to other contexts. Additionally, educators may struggle to evaluate student performance when AI tools are used extensively \cite{mhlanga2023open}. Therefore, it is essential to develop effective ways that can help measure student knowledge and skills while reducing their reliance on such AI tools during assessments. 

As per the literature, Bloom's taxonomy is a widely accepted framework for developing educational objectives and assessing learning outcomes. The framework categorises cognitive skills into six levels, including remembering, understanding, applying, analysing, evaluating, and creating \cite{forehand2010bloom, krathwohl2002revision}. The use of Bloom's taxonomy in designing effective assessment tools can help measure the depth of student understanding and knowledge. To the author's knowledge, no studies have been conducted on optimising the best Bloom's Taxonomy keywords that can help reduce the reliance on AI tools. 

To tackle the aforementioned challenges, this paper proposes an evolutionary approach to optimize Bloom's Taxonomy keyword selection for generating questions that generative text AI tools have low confidence in answering. The approach is expected to encourage students to think critically and engage with the material more deeply, ultimately improving their learning experience. In this paper, a genetic algorithm (GA) is used to evolve a set of chromosomes (solutions), each of which represents a number of Bloom's Taxonomy keywords. Those keywords are used to generate a new version of the question of interest using the generative text AI tool. Then, the confidence of the generative text AI tool in answering the new question represents the quality of the solution. The algorithm iterates through selection, crossover, and mutation processes to evolve the solutions over a fixed number of generations. The evolutionary process continues to minimise such a confidence value. 

The algorithm is evaluated on a number of questions taken from the Data Structures and Representation (DSR) course being taught at the University of New South Wales Canberra, Australia. The results demonstrate that the proposed approach is able to identify a set of keywords that can be used to generate questions that reduce ChatGPT's confidence in answering them from an average of 95\% to 45\% and even generate questions with 0\% confidence in their answers. This reduction in confidence is expected to result in assessments that are less susceptible to AI-generated answers, ultimately leading to better evaluation of student understanding and knowledge.

It is important to highlight that a brute force approach (trying all combinations of Bloom's Taxonomy keywords) is not efficient as the generative text AI tool can restructure the question in many different ways, making the search space huge.

The rest of the paper is organized as follows. Section \ref{G_AI_tools} provides an overview of some AI tools and related topics covered in this paper. Section \ref{proposed approach} describes the proposed approach. Section \ref{sec:results} discusses the results obtained. Finally, Section \ref{conclusion} concludes the paper and discusses future work.

\section{Overview} \label{lit rev}
This section provides a brief overview of generative AI Tools, Bloom's Taxonomy and evolutionary algorithms.

\subsection{Generative AI Tools} \label{G_AI_tools}

Recently, the domain of text generation powered by AI has witnessed substantial growth due to advancements in machine learning and deep learning methodologies. These tools have been employed across various applications, such as natural language processing (NLP), content creation, and automated interactions \cite{brown2020language}. Below is a summary of some of those tools.

\subsubsection{OpenAI's GPT-3}
The Generative Pre-trained Transformer 3 (GPT-3), devised by OpenAI, is an autoregressive language model containing a notable 175 billion parameters \cite{brown2020language}. GPT-3 demonstrates exceptional skill in producing text that mimics human-like writing, handling translation-related tasks, and answering various questions. The thorough pre-training process employed by GPT-3 allows for the creation of contextually relevant and coherent output.
\subsubsection{Google AI's BERT}
The Bidirectional Encoder Representations from Transformers (BERT) is an AI model developed by Google AI, designed to excel in natural language understanding tasks \cite{devlin2018bert}. BERT's masked language model approach allows for the consideration of word context from both directions, thereby improving performance in question-answering tasks.

\subsubsection{Google AI's T5}

In T5, a method was developed that consolidates various NLP tasks into a single text-based format, wherein both inputs and outputs are represented as textual strings. This is in contrast to BERT-inspired models that only generate class labels or specific input segments. Utilizing this text-oriented framework, the identical model, loss function, and hyperparameters can be employed across a wide range of NLP tasks, including machine translation, condensing documents, answering questions, and categorizing tasks such as sentiment analysis. Furthermore, T5 can be tailored for regression tasks by training it to forecast a number's textual form rather than the numeric value itself \cite{roberts2020t5}.

This paper focuses only on GPT-3 due to its popularity and performance.

\subsection{Bloom's Taxonomy} \label{BT}

Bloom's Taxonomy is a hierarchical classification of cognitive learning objectives. It serves as a framework for educators to develop learning objectives, design educational materials, and assess students' progress in a systematic and structured manner \cite{bloom1956handbook}. Bloom's Taxonomy consists of six levels, ranging from basic knowledge to higher-order thinking skills. Krathwohl \cite{krathwohl2002revision} introduced a revised version of Bloom's Taxonomy framework. The new framework had two dimensions: Knowledge and Cognitive Processes. The Knowledge dimension closely resembled the original taxonomy's subcategories, while the Cognitive Processes dimension revised category names: Remember (previously Knowledge), Understand (previously Comprehension), Apply, Analyze, Evaluate, and Create (previously Synthesis and now the top category). Some verbs that belong to each category are shown below.
\begin{itemize}
\item Remembering: recognize, recall, retrieve, reproduce, list, name, define, identify, match, label, select, outline
\item Understanding: explain, summarize, paraphrase, infer, classify, compare, contrast, exemplify, generalize, predict, discuss
\item Applying: apply, use, demonstrate, illustrate, interpret, operate, schedule, sketch, solve, modify, relate, choose
\item Analyzing: analyze, distinguish, categorize, diagram, differentiate, discriminate, infer, select
\item Evaluating: evaluate, appraise, criticize, judge, justify, support, weigh, assess, interpret, argue, compare, contrast, rate
\item Creating: create, design, invent, compose, plan, formulate, generate, hypothesize, produce, develop, originate, arrange, construct
\end{itemize}

\subsection{Evolutionary Algorithms}

Evolutionary algorithms (EAs) have been widely used to tackle complex problems by imitating the natural evolution process. There are different EAs, including GAs, differential evolution (DE), evolution strategies (ES), and evolutionary programming (EP) \cite{elsayed2012evolutionary, eiben2015introduction}. 

Although these EAs share some common steps, they vary in the order of these steps and how they create an initial group of possible solutions (population). Each possible solution can be represented in different formats, such as real numbers, integers, or strings. The idea of natural selection ensures that the best solutions have a higher chance of influencing future generations.

To create new solutions, EAs use two main processes called recombination (crossover) and mutation. Recombination combines selected solutions to produce new ones, while mutation introduces small changes to a single solution to maintain variety. These new solutions may compete with their parents, with the best ones being chosen to form a new population, or all new solutions might be selected, keeping some of the best from the previous population.

These steps are repeated until a satisfactory solution is found or certain stopping conditions are met. By using this nature-inspired approach, EAs offer a powerful way to solve complex problems in a variety of fields.

\section{proposed approach} \label{proposed approach}
This section describes the proposed framework and its components.

\subsection{General Framework} \label{general framework}

As outlined in Algorithm \ref{alg:evolutionary}, the proposed approach begins by generating $p$ solutions, also referred to as chromosomes, each consisting of $n$ randomly chosen keywords from Bloom's Taxonomy. To optimize the number of keywords used in the solution, empty words are added to the list, which means each chromosome can have $i$ keywords, where $i \in 1, 2, ..., n$. Subsequently, the quality of each chromosome is calculated, as described in Section \ref{fitness function}.

Aiming to minimize the confidence score of answers generated by ChatGPT, the solutions undergo evolution through the selection, crossover, and mutation operators, as detailed in Sections \ref{selection}, \ref{crossover}, and \ref{mutation}. The offspring solutions created through these genetic operators are then evaluated based on their fitness and selected for the next generation, with the best solution in the population always surviving to the next generation (a process known as elitism).

The evolutionary process continues until the maximum number of generations is reached, with the objective of generating a high-quality solution to the original question using Bloom's Taxonomy as guidance.

\begin{algorithm}[]
  \DontPrintSemicolon
  \KwIn{Bloom's Taxonomy keywords, $p$ (number of solutions), $n$ number of genes, $g_{max}$ maximum number of generations}
  \KwOut{Optimised set of blooms taxonomy keywords}

  generate $p$ initial solutions, each consisting of $n$ randomly selected keywords from Bloom's Taxonomy\;
  
  calculate the quality of each chromosome using the fitness function (section \ref{fitness function})\;
  
  \While{not at $g_{max}$}{
    select parents using the tournament selection\;
    generate offspring using the crossover and mutation operators\;
    evaluate the fitness of the offspring solutions\;
    set the new offspring as the new population\;
    replace the worst solution in the new population with the best solution in the previous generation  (elitism)\;
  }
  \caption{Proposed Approach}
  \label{alg:evolutionary}
\end{algorithm}


\subsection{Fitness Evaluation} \label{fitness function}
To evaluate the quality of each solution, the following steps are performed:
\begin{itemize}
\item OpenAI's GPT-3 model\footnote{this is done by using the OpenAI API with the GPT-3 model variant "text-davinci-003" used} is utilized to generate a question that incorporates the keywords in the chromosome. This is achieved by asking the model "\textit{rephrase this question to use keywords in the chromosome}."
\item OpenAI's GPT-3 model is used to answer the newly generated question, by asking "\textit{what is the answer for the generated question?}"
\item OpenAI's GPT-3 model is employed to provide a confidence score from 0 to 1 for the generated answer, by asking "\textit{give me a confidence score between 0-1 for your answer. Just write the number without any text}."
\end{itemize}

This final confidence score represents an estimate of the likelihood of the tool's response accuracy based on how closely it matches the patterns in the training data. As a result, this confidence score is used as the fitness value of the chromosome that the algorithm aims to minimize.

\subsection{Selection} \label{selection}
Tournament selection is used, which involves randomly selecting two individuals from the population. The winner, based on its fitness value, is then copied into the mating pool. This process continues until $p$ solutions are added to the mating pool.

\subsection{Crossover} \label{crossover}
For each chromosome in the mating pool, another solution is generated. In this process, for each gene position in the chromosome, a probability of $cr=0.5$ is used to decide whether to take the gene from this chromosome or from another randomly chosen parent in the mating pool, keeping in mind that no duplicate genes can exist in a single chromosome.

\subsection{Mutation} \label{mutation}
To maintain diversity, a mutation process takes place for each solution generated in the crossover process. For each gene, a probability of $mr$ is used to decide whether to take the gene from the chromosome generated by the crossover process or randomly select it from Bloom's Taxonomy keywords.

\section{Experimental Results}
This section discusses the results and analysis obtained from this study.

\subsection{Experimental Setup}
To evaluate the effectiveness of the proposed approach, a series of experiments were conducted. A set of questions taken from DSR exams at UNSW Canberra were used. These questions can be found in the appendix. Additionally, the following Bloom's Taxonomy keywords were selected as they are suitable for a programming course.

In the research paper, the Bloom's Taxonomy keywords used as input are as follows:

\begin{itemize}
\item \textbf{Knowledge:} define, describe, identify, list, recall, recognize, state
\item \textbf{Comprehension:} classify, explain, interpret, summarize, translate
\item \textbf{Application:} apply, demonstrate, implement, use
\item \textbf{Analysis:} analyze, compare, contrast, differentiate, examine, test
\item \textbf{Evaluation:} assess, evaluate, judge
\item \textbf{Creation:} design, develop, plan
\end{itemize}

These keywords represent various cognitive levels and skills, ranging from basic knowledge recall to higher-order thinking skills such as analysis, evaluation, and creation. These words also align with a DSR programming course. It is worth noting that the algorithm is flexible enough to accept more words. 

The other parameters for the algorithm are $p = 20$, $n=3$, mutation rate ($mr$) = 0.1, crossover rate ($Cr$) = 0.5, the maximum number of generations ($g_{max}$) = 10, and elitism size = 1. The algorithm was run three times for each question. Additionally, the configurations used for the OpenAI engine are engine: \texttt{text-davinci-003}, temperature: $\tau = 0.7$, stop token: None, and the number of responses: $r = 1$.

It is important to mention that the maximum chromosome size is set to 3 to avoid creating overly complicated questions. Also, the number of runs and generations is small, due to the limited usage OpenAI allows each month (USD120). However, efforts will be made to mitigate this in the future.

\subsection{Results and Discussion} \label{sec:results}

\begin{table*}[]
\caption{Results of the Proposed Algorithm}
\label{tab:results}
\centering
{\begin{tabular}{|l|ll|l|}
\hline
\multirow{2}{*}{Questions} & \multicolumn{2}{l|}{Cofidence Score}             & \multirow{2}{*}{Best Chromosome(s)}                                                                                                                                    \\ \cline{2-3}
                           & \multicolumn{1}{l|}{Before} & After Optimisation &                                                                                                                                                                      \\ \hline
Q1                         & \multicolumn{1}{l|}{0.9}    & 0.8                & {[}'interpret', 'contrast',   'test'{]}                                                                                                                              \\ \hline
Q2                         & \multicolumn{1}{l|}{0.95}   & 0.5                & \begin{tabular}[c]{@{}l@{}}{[}'describe', 'demonstrate', 'state'{]}\\ {[}'implement', 'analyze',   'recall'{]}\\ {[}'evaluate', 'interpret', 'state'{]}\end{tabular} \\ \hline
Q3                         & \multicolumn{1}{l|}{0.85}   & 0                  & {[}'describe', 'list', 'examine'{]}                                                                                                                                  \\ \hline
Q4                         & \multicolumn{1}{l|}{0.95}   & 0.5                & \begin{tabular}[c]{@{}l@{}}{[}'  ',   'analyze', 'design'{]}\\ {[}'    ', 'develop', 'differentiate'{]}\end{tabular}                                                 \\ \hline
Q5                         & \multicolumn{1}{l|}{0.95}   & 0                  & {[}'examine', 'contrast', 'evaluate'{]}                                                                                                                              \\ \hline
Q6                         & \multicolumn{1}{l|}{0.95}   & 0.5                & {[}'implement', 'evaluate', 'compare'{]}                                                                                                                             \\ \hline
Q7                         & \multicolumn{1}{l|}{0.95}   & 0.5                & {[}'analyze', 'identify', 'judge'{]}                                                                                                                                 \\ \hline
Q8                         & \multicolumn{1}{l|}{0.9}    & 0.8                & \begin{tabular}[c]{@{}l@{}}{[}'assess', 'explain',   'interpret'{]}\\ {[}'develop', 'interpret', 'apply'{]}\\ {[}'recognize', 'compare', 'translate'{]}\end{tabular} \\ \hline
\end{tabular}}

\end{table*}

As mentioned earlier, the primary goal of the algorithm is to find the best set of keywords to generate questions that reduce GPT-3's confidence in answering them. A lower confidence score signifies improved performance. The results are presented in Table \ref{tab:results}, which illustrates the confidence scores of GPT-3 before and after optimization, along with the best chromosome(s) found for each question.

Upon analyzing the results, it is evident that the proposed algorithm successfully reduces the confidence scores of ChatGPT for all the questions in the dataset. This demonstrates the effectiveness of the algorithm in achieving its primary goal.

For instance, Question 1 (Q1) exhibits a reduction in the confidence score from 0.9 to 0.8. Similarly, Q2 shows a more significant decrease in confidence score from 0.95 to 0.5, indicating a substantial improvement in the performance of the algorithm.

The most notable result is observed for Q3, where the algorithm managed to completely eliminate the model's confidence in its answer, reducing the score from 0.85 to 0. This indicates that the algorithm was highly effective in guiding ChatGPT towards producing more uncertain responses for this particular question. The same occurred in Q5. The remaining questions (Q4, Q6-Q8) also show a consistent trend of reduced confidence scores after optimization, and for some of them, the algorithm managed to find multiple optimal solutions.

In addition to the reduction in confidence scores, the best Bloom's Taxonomy keywords for each question were analyzed. For Q1, the best chromosome found by the algorithm includes the keywords 'interpret', 'contrast', and 'test'. These keywords appear to be well-suited to the question, as they capture the essence of the underlying cognitive processes involved in understanding the problem and analyzing the information provided.

In the case of Q2, the algorithm found multiple optimal solutions, including keywords such as 'describe', 'demonstrate', 'state', 'implement', 'analyze', 'recall', 'evaluate', and 'interpret'. These keywords cover a range of cognitive processes, from basic recall to higher-level evaluation and interpretation, which suggests that the algorithm is capable of identifying diverse combinations of keywords that can effectively reduce ChatGPT's confidence in its responses.

For Q3, the algorithm selected the keywords 'describe', 'list', and 'examine'. These keywords emphasize the need to provide detailed information and conduct a thorough examination of the problem, which may have contributed to the complete elimination of ChatGPT's confidence in its answer to this question. For Q5, the keywords ('examine', 'contrast', and 'evaluate') come from high-level (analyze and evaluate) categories.

For Q4, which is about writing a Java method, the algorithm selected two words ([analyze and design] or [develop and differentiate]), one from Bloom's level 4 and another from Bloom's level 6. This finding may help educators in structuring their questions in programming assignments.

The best chromosomes for the remaining questions (Q6-Q8) also contain combinations of Bloom's Taxonomy keywords. For Q6, the best words came from levels 3-5, while Q7 contains words from levels 1 and 5.

\subsection{Validation Step}
The aim of this step is to measure the correctness of the answers generated by GPT-3 for the newly generated questions. As the course convener, I assessed the answer 
for each newly generated best question, with the following analysis summarized:

\begin{itemize}
\item Q1: Although the tool had still a high confidence value, the answer is incorrect, as the AI tool selects the wrong Java collection for the given scenario. 
\item Q2: The answers are mixed, i.e., some being correct ($O(n)$) and others being wrong ($O(n^2)$).
\item Q3: The answer is either wrong or incomplete. The model provides a generic response without fully addressing the question.
\item Q4: The answer is incorrect, as it does not design or analyze the code - it simply writes the code, which is about 90\% correct.
\item Q5: The answer is incorrect, as the steps the AI tool provides are incomplete, and the final array [19, 8, 41, 72, 50, 60] is incorrect.
\item Q6: The answer is partially correct, but it does not fully address the question.
\item Q7: The answer is incorrect and provides very generic information with some inaccuracies.
\item Q8: The answer is good but incomplete, as it does not fully address all aspects of the question.
\end{itemize}

In summary, the AI model struggles to provide accurate and complete answers to the newly generated questions. This analysis highlights the model's limitations and emphasizes the importance of using different keyword combinations that span multiple cognitive levels to effectively reduce the reliance on AI-generated answers.

\section{Conclusion and Future Work} \label{conclusion}

In conclusion, this research proposed an evolutionary algorithm that aims to identify the best set of Bloom's Taxonomy keywords to generate programming questions that AI tools, such as ChatGPT, have low confidence in answering. The analysis focused on the model's performance in addressing questions across multiple cognitive levels, as defined by Bloom's Taxonomy, thereby highlighting the limitations of the AI model in providing accurate and complete answers.

The results demonstrated that a careful selection of keywords from different cognitive levels can significantly challenge the AI model's capabilities. Moreover, the optimization algorithm proved to be effective in reducing the confidence scores of ChatGPT, indicating that incorporating diverse keywords in the question design can further impact the performance of the AI model.

This study emphasizes the importance of providing educators with the necessary knowledge and tools to design questions that not only assess students' understanding but also minimize their dependency on AI-generated answers. By comprehending the limitations of AI models like ChatGPT and strategically crafting questions, educators can facilitate active engagement in the learning process, promoting the development of critical thinking and problem-solving skills in students. Overall, this research contributes to the growing body of knowledge on the application of AI in education and offers valuable insights for both educators and AI researchers.

In future research, I plan to increase the number of generations and runs of the optimization algorithm to improve the quality of the generated questions and further reduce the confidence score of ChatGPT. Mitigating the challenge of limited monthly usage and costs associated with OpenAI API is also crucial, as it could potentially affect the scalability of the approach. Additionally, I aim to expand the scope of the analysis to include more questions and cover a wider range of topics and courses. Finally, the objective function used to evaluate solutions can be improved by considering the uncertainty of the generated solutions by the tools.

\bibliographystyle{IEEEtran}
\bibliography{refs}

\small
\appendix
\section{Sample Questions}
\label{sec:sample-questions}

The following are the questions used to evaluate the algorithm. These questions are taken from a final-year exam for the DSR course.

\begin{enumerate}
\item You are designing a Java program to monitor a network of IoT (Internet of Things) devices. The program will receive a continuous stream of data from various devices, including temperature, humidity, and air quality readings. The program should analyze the data in real-time and raise an alert if any of the readings exceed a certain threshold. The alerts should be sent to a centralized monitoring system, and a log of all the data should be kept for historical analysis. The program should be able to handle a large number of devices and scale accordingly as the network grows. Just indicate which Java Collection type would be most appropriate to use in this program.
\item State the Big-O order of the time complexity for the following code fragments and briefly describe how you determined its value: \texttt{i = N*2; while (i>0) \{ i = i-2; sum += i; \}}

\item Consider the following hash function for string-valued keys of length at least 6, for a hashtable of size 1024. \texttt{int hash( String s ) \{ return (s.charAt(0) +  s.charAt(1) - s.charAt(2) + s.charAt(3) - s.charAt(4))  \% 1024; \}} State whether you believe this hash function is likely to be effective or not, when keys are the course codes (such as we use at UNSW Canberra) for the university courses? Justify your opinion in a few sentences.

\item A method is required that will write out data from two collections to a specified file. The two collections represent a group of scores (Integer objects), and the names of people who achieved those scores (Strings). That is, the first element from the scores collection matches the first element from the names collection; etc. The data is to be output in plain ASCII format (to the file) with one score and associated name per line, with a colon and blank space placed between the score and the name. The signature of the method is: \texttt{public static boolean write2File(String fname, Set<String> names, List<Integer> scores)}. The method is to return true if it was able to output the data, and false otherwise. Write the method.

\item Show every step of the (ascending) heapsort algorithm (performed in place) on the following array of numbers [60, 19, 41, 72, 50, 8].

\item The array-based implementation for a binary search tree can be done either by a computational strategy or a simulated link strategy. Using the following computational strategy-based implementation of a tree: E C H A D. a), In what order will a pre-order traversal visit the nodes in the tree, assuming the left child is visited before the right?

\item Assume a list is implemented using a singly linked list with a head pointer. Write a generic add method that shows how to add an element at a specific location. You may assume the existence of the following variables and information: \begin{itemize}
    \item headNode – pointer to the head node  
    \item position – location at which the new node is added 
    \item Node class that represents a node in the singly linked list with two pointers 
    \item nextNode - pointer to the next node 
    \item dataElement - a generic data element it holds
\end{itemize}

\item Linked lists can be implemented in a number of variants, using single or double links, and with a head pointer only or both head and tail pointers. Describe and justify the most efficient linked list implementation when a basic queue data structure is needed.
\end{enumerate}

\end{document}